\documentclass[12pt,preprint]{aastex}

\usepackage{emulateapj5}
\usepackage{onecolfloat5}
\usepackage{apjfonts}

\def\plotone#1{\centering \leavevmode
\includegraphics[clip=, width=.85\columnwidth]{#1}}
\def\plottwo#1#2{\centering \leavevmode
\includegraphics[width=.45\columnwidth]{#1} \hfil
\includegraphics[width=.45\columnwidth]{#2}}
\def\plotthree#1#2#3{\centering \leavevmode
\includegraphics[width=.3\columnwidth]{#1} \hfil
\includegraphics[width=.3\columnwidth]{#2} \hfil
\includegraphics[width=.3\columnwidth]{#3}}

\newcommand{\spa}{\mbox{ }}
\def\gsim{\;\rlap{\lower 2.5pt
 \hbox{$\sim$}}\raise 1.5pt\hbox{$>$}\;}
\def\lsim{\;\rlap{\lower 2.5pt
   \hbox{$\sim$}}\raise 1.5pt\hbox{$<$}\;}

\begin{document}

\twocolumn[

\title{Can We Probe the Atmospheric Composition of an Extrasolar Planet from\\ its Reflection Spectrum in a High--Magnification Microlensing Event?}

\author{David S. Spiegel\altaffilmark{1}, Michel Zamojski\altaffilmark{1}, Alan Gersch\altaffilmark{1,2}, Jennifer Donovan\altaffilmark{1}, Zolt\'an Haiman\altaffilmark{1}}

\affil{$^1$Department of Astronomy, Columbia University, 550 West 120th Street, New York, NY 10027}

\affil{$^2$Depertment of Astronomy, University of Maryland, College Park, MD 20742}

\vspace{0.5\baselineskip}

\email{dave@astro.columbia.edu, michel@astro.columbia.edu, agersch@astro.umd.edu, jen@astro.columbia.edu, zoltan@astro.columbia.edu}

\begin{abstract}
We revisit the possibility of detecting an extrasolar planet
around a background star as it crosses the fold caustic of a
foreground binary lens.  During such an event, the planet's flux can
be magnified by a factor of $\sim 100$ or more.  The
detectability of the planet depends strongly on the orientation of its
orbit relative to the caustic.  If the source star is inside the
inter--caustic region, detecting the caustic--crossing planet is
difficult against the magnified flux of its parent star.  In the more
favorable configuration, when the star is outside the inter--caustic
region when the planet crosses the caustic, a close--in
Jupiter--like planet around a Sun--like star at a distance of 8 kpc is
detectable in 8-minute integrations with a 10m telescope at maximal
$S/N\sim15$ for phase angle $\phi\sim10\degr$.  
In this example, we find further that the presence of methane, at its measured
abundance in Jupiter, and/or water, sodium and potassium, at the abundances
expected in theoretical atmosphere models of close--in Jupiters, could be
inferred from a non--detection of the planet in strong broad absorption bands
at $0.6-1.4\mu$m caused by these compounds, accompanied by a $S/N\sim 10$
detection in adjacent bands.  We conclude that future generations of large
telescopes might be able to probe the composition of the atmospheres of distant
extrasolar planets.

\end{abstract}
]

\section{Introduction}
\label{intro}

With the first discovery a dozen years ago of a planet orbiting a star
other than our Sun (Wolszczan \& Frail, 1992), astronomy finally
entered an age in which we could hope to answer scientific questions
about distant planets, with the ultimate aim of detecting and
characterizing an extrasolar Earth-like planet.  Among the most
tantalizing questions is: What is the chemical composition of an
extrasolar planet?
In this paper, we suggest a way to test for the presence of certain
compounds in the atmospheres of extrasolar planets at much greater
distances than has previously been discussed.

While the vast majority of the $\sim 155$ currently known extrasolar
planets have been discovered in radial velocity surveys (e.g., see
Marcy \& Butler, 1998 or Woolf \& Angel, 1998 for reviews
\footnote{For up--to--date information on the current status of these
searches, see \textsl{http://www.obspm.fr/encycl/encycl.html} and
\textsl{http://www.exoplanets.org}.}), two methods have recently been
proposed to search for extrasolar planets via their gravitational
microlensing signatures.  These methods are complementary to the
radial velocity surveys, in that they can detect planets at larger
distances, well beyond our solar neighborhood, and one of these
methods has the advantage of potentially providing information about
the spectrum and therefore the composition of the planet (transit
surveys have the same two advantages; e.g. Charbonneau et al. 2002).

Mao \& Paczynski (1991) and Gould \& Loeb (1992) suggest that as a
background star passes behind a lens--star with a companion planet,
the planet could be detected as lens, since it will cause a secondary,
sharp spike in the source star's light--curve.  Indeed, two years ago,
a lens-plane planet was finally discovered (Bond et al., 2004).
Recently, Graff \& Gaudi (2000, hereafter GG00)
and Lewis \& Ibata (2000) have suggested that
extrasolar planets might instead be detected in the source-plane, as they
cross the caustics of a foreground lens system and are highly magnified
relative to their parent star (Heyrovsky \& Loeb 1997 also discuss
from a theoretical perspective the possibility of using microlensing 
light--curves to probe structures in the source plane, in their case
the structure of a background star behind a point--like lens, and 
other authors since then have carried out such studies, e.g. Albrow et
al., 1999, Castro et al., 2001).

While detecting the planet as a lens, as Mao \& Paczynski suggest,
has the potential to reveal a statistically important sample of
extrasolar planets, the drawback is that we receive no information
about the planets except for perhaps their masses and projected
separations from their host stars.  Reflected light from a planet,
however, contains information about physical parameters of the planet
(presence and sizes of rings, satellites, spots and bands, for
example).  Detecting a planet as a lensed source therefore holds the promise
of allowing these parameters to be measured, as suggested by Gaudi, Chang, \&
Han (2003, hereafter GCH03; Lewis \& Ibata 2000 suggest further that
polarization fluctuations during microlensing events could be indicative of
properties of planetary atmospheres).  In the present work, we investigate the
viability of detecting an extrasolar planet as a microlensed source,
and the extent to which a measurement of the magnified reflection spectrum can 
be used to glean information about the planet's atmospheric composition.

An unperturbed, isolated point--like lens (such as a single planet or a star)
produces a point--like caustic.  A binary lens, however, can produce a closed
caustic curve, consisting of a set of piecewise concave curves that
meet in cusps.  In the present context, a binary lens, then, has several
advantages over a point--lens: first, the relatively large spatial extent
(compared to a point) of the binary lens caustic implies a much larger region in
the sky in which for a high--magnification event to occur; second, since the
caustic of a binary lens is a closed curve, caustic crossings come in pairs, and
the second crossing can be anticipated; third, both star and planet can cross the
caustic of a binary lens, while it is unlikely that both would cross the
point--caustic of a point lens.  When a background star with a companion planet
crosses the caustic of a binary lens, a unique observational signature will
be produced in the light--curve.  If such a signature is detected on ingress
(or, if the lensed light--curve shows, at least, that the
star has entered the inter--caustic region of a binary lens),
GG00 suggest that many observatories
could train their telescopes on this system so as to obtain dense
sampling of the light--curve at egress (exiting the caustic region).
If the planet's reflected light is sufficiently magnified,
multi--color light--curves, or even detailed time-dependent spectra
might, in principle, be obtained.  Such spectral binning of the signal
would shed light on the wavelength-dependence of the planet's albedo,
which could in turn yield information about the chemical composition
of the planet's atmosphere.

GCH03 suggest that morphological features such as moons or rings around
extrasolar planets may be detectable, and they find a signal-to-noise ratio of
$\sim 15$ for $I$-band detection of a planet in a typical
planet-star-lens configuration with a 10m telescope.  If the light (in
a given wavelength range) is split up into $N$ bands, the
signal-to-noise ratio should go down roughly as $1/\sqrt{N}$.
Signal-to-noise is also directly proportional to the diameter of the
telescope's aperture.  This suggests that with a 10m--class telescope,
light could be split up into a few broad spectral bins before
signal-to-noise becomes unacceptably low, and motivates us to examine
whether useful information about the atmospheric composition of the
planet could be obtained with this method.

The rest of this paper is organized as follows. In \S~\ref{model}, we
present our model of a planetary caustic--crossing, including a
detailed discussion of both the model of the planet and the
computation of the caustic--crossing light--curve. In
\S~\ref{detectability}, we discuss the detectability of extrasolar
planets through microlensing. In \S~\ref{prspectra}, we describe the
albedos and reflection spectra of gas--giant planets in our own solar
system. In \S~\ref{specres}, we analyze the possibility of determining
the wavelength--dependence of the albedo of a microlensed extrasolar
planet.  In \S~\ref{discussion} we present a detailed discussion of the
factors that affect the S/N of the detection of planets with various
features in their reflection spectra. 
Finally, in \S~\ref{conc}, we discuss the limitations of
current technology, and conclude with projections of what may be
possible with future instruments.

\section{Modeling Planetary Caustic--Crossing Events}
\label{model}

\subsection{The Planet--Star System}
\label{pssystem}

\begin{figure}[ht]
\plotone{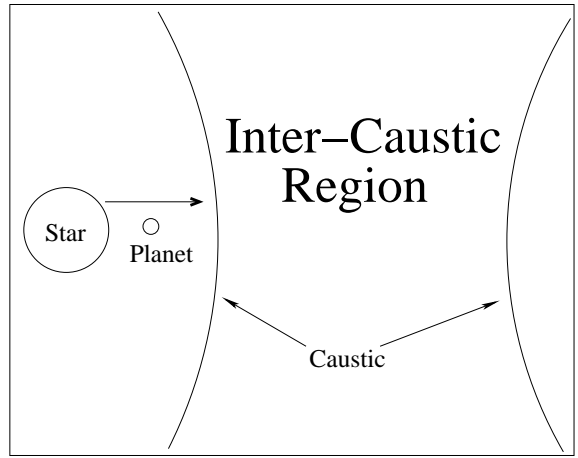}
\caption{Schematic illustration of the planet--star system in
the ``planet--leading'' configuration.  The planet--star system is
moving to the right, while the caustic stays still.  In the reverse
(``planet-trailing'') configuration, the planet would be to the left
of the star as they move to the right.}
\label{fig:intercaustic} 
\end{figure}

 We consider a star with a companion planet as it crosses the fold--caustic
of a binary lens.
Figure \ref{fig:intercaustic} shows an illustration of the configuration we
model.
The observed surface--brightness of the planet at a given wavelength
depends on properties of the star, the planet, and their relative
geometry -- specifically, the stellar flux, the albedo and phase of 
the planet, and the reflection or scattering properties of its atmosphere.  
For the stellar spectrum, we adopt that of a G0V star, and for the
wavelength--dependent albedo of the planet's atmosphere, we use the
gas giants in our own solar system as a guide (both to be described 
in more detail in \S~\ref{prspectra} below, where spectral features 
are considered).  The planetary phase is given by the angle $\phi$ between the
line--of--sight and the ray from the star to the planet (e.g. $\phi=0\degr$
corresponds to the ``full-moon'' phase), as described in, e.g., GG00, GCH03,
and Ashton \& Lewis (2001).  GCH03 adopted and compared two simple
reflectance models (uniform and Lambert scattering) that prescribe the angular
dependence of reflectivity; and Ashton \& Lewis (2001) considered the effects
of planetary phase.  Neither of these studies, however, considered
simultaneously the effects of the planet's phase and its reflectance model
on the lensed light--curve.  In our studies, we compared three different
reflectance models: uniform, Lambert, and Lommel-Seeliger reflection (see,
e.g., Efford, 1991 for more detailed discussions of these models, and see
the Appendix for details on the computation of planet-models).

\begin{figure}[ht]
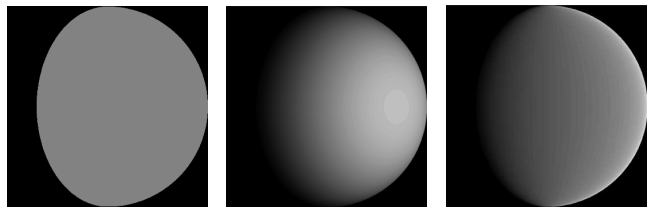

\plotthree{fig2a.epsf}{fig2b.epsf}{fig2c.epsf}
\caption{Illustration of the surface brightness of planets illuminated
by a star at a phase angle of $\phi=45$ degrees.  The three panels
assume different reflectance models. {\em Left panel:} uniform
illumination; {\em Middle panel:} Lambert reflectance; {\em Right
panel:} Lommel--Seeliger reflectance.  The color maps differ slightly
between the different models; all three scatter approximately equal
amounts of light toward the viewer.}
\label{fig:planets} 
\end{figure}

In Figure~\ref{fig:planets}, we illustrate the surface brightness maps
of three planets, one for each reflectance model described, each at
fixed phase $\phi=45\degr$.  The maps were created numerically on a
square grid of $401\times401$ pixels that we find to be sufficiently
fine to converge on the light--curves we obtain below.  We concur with
GCH03 that with current technology it would be impossible to infer the
true reflectance model of an extrasolar planet during a microlensing
event, and so we use only one model, Lommel-Seeliger reflectance -- 
which we expect to be the most realistic one, in calculating the
light--curves that we present below.

\subsection{Modeling the Caustic--Crossing Event}
\label{ccevent}
For a good description of the details of gravitational microlensing,
including see, for example, Mao \& Paczynski (1991); for details
regarding the generation and shape of fold-caustics, see GG00.  We
assume that the planet--star system described in \S~\ref{pssystem}
above is in the source plane of a binary lens.  The lensing stars are
massive enough and close enough to one another that they generate a
fold--caustic in the source-plane, a closed curve of formally infinite
magnification.  The caustic is considered to be a straight line
(we follow GG00 and GCH03 and assume a low probability of crossing
the caustic near a cusp) that sweeps across the planet and star.  
We assume that the plane of the planet's orbit is edge--on and is
normal to the caustic at the point where the star-planet system crosses;
we will argue in \S~\ref{specres} below that this simplification is
not critical to our results.  A source is magnified by the binary lens
proportionally to the inverse square-root of the source's distance from
the caustic when it is in the inter--caustic region (ICR) and not
otherwise.  We use the discretized magnification equation given by
Lewis \& Belle (1998) and Ashton \& Lewis (2001):
\begin{equation}
\label{ampd}
\mathcal{A}_{\rm{pix}}(x_k) = 2 \frac{\kappa}{\Delta x} \left(
        \sqrt{x_k + \Delta x} - \sqrt{x_k} \right) + \mathcal{A}_0,
\end{equation}
where $\mathcal{A}_{\rm{pix}}$ is the magnification of the pixel,
$\kappa$ is a constant close to unity that represents the
``strength'' of the caustic, $\Delta x$ is the width of a single
pixel, and $x_k$ is the distance of the $k$th pixel from the caustic.
In equations~(\ref{ampd}), distances are measured in units of the
Einstein radius of the lens system:
$\theta_E = \sqrt{2 R_{\rm{Sch}}/D}$, where $R_{\rm{Sch}} = 2 G M/c^2$
is the Schwarzschild radius of the lens, and
$D \equiv D_{\rm{os}} D_{\rm{ol}}/D_{\rm{ls}}$ ($D_{\rm{os}}$, $D_{\rm{ol}}$,
and $D_{\rm{ls}}$ are the distances between the observer and source, the
observer and lens, and the lens and source, respectively).  In our
model, we use equation~(\ref{ampd}) to compute the brightness of each
pixel across the face of the planet.  We then sum the contributions
from all of the $401\times401$ pixels to determine the total
brightness at a given position $x$ (corresponding to a given time
during the lensing event).

\subsection{Preview of Results}
\label{preview}
Using the model described above, we study a number of different
scenarios. In all cases, we assume that the source star is a clone of
HD209458, a G0V 8 kpc away with a companion planet that has the
properties of HD209458b, i.e. a ``Hot Jupiter,'' with radius $R_p =
1.35$ times the radius of Jupiter, and with orbital radius $a = 0.046$
AU (for details of the discovery of HD209458b, see Henry et al. 2000;
Charbonneau et al. 2000).  In order to reproduce the published results
of GCH03, we assume generous viewing conditions with albedo $A = 1$
(all incident light is reflected).  To model a more realistic
situation, we examine several other albedo models, including Jupiter's
albedo and a ``gray atmosphere'' -- a constant,
wavelength--independent albedo.  The lens stars are assumed to be
typical bulge stars, 6~kpc away, each with a mass of $0.3{\rm M_\odot}$.

We first consider a simple estimate of the flux from such a system.
The un--magnified flux from a solar-type star 8 kpc away is quite low;
$F_* = L_\sun / 4 \pi (8 \rm{kpc})^2 \approx 5 \times 10^{-13}~\rm{erg}~\rm{cm}^{-2}~\rm{s}^{-1}$. Since a typical photon
($\sim 500$ nm) from such a star carries about $4 \times 10^{-12}$
ergs, this flux corresponds to a photon number flux of approximately
$0.1 \rm{photons}$ $\rm{cm}^{-2}$ $\rm{s}^{-1}$.  Even a large,
close--in planet, such as the one under consideration, subtends a
small solid angle from the star's perspective; so even with an albedo
of $A=1$, the flux of photons from the planet is reduced by a factor
of $\gtrsim 10^4$ from the stellar flux. As a result, the flux from
the planet is $\approx 6 \times 10^{-6}$ photons $\rm{cm}^{-2}$
$\rm{s}^{-1}$.  This is the well--known reason why gravitational
microlensing is essential for detecting the reflected light of a
planet around such a distant star.

Crossing a fold-caustic can lead to impressive magnification.  In the
situation under consideration, the Einstein radius of the lens is
approximately 4000 times the radius of the planet, which means that,
according to equation (\ref{ampd}), a magnification factor of
$\mathcal{A} \sim \sqrt{4000} \approx 60$ can be achieved.\footnote{Our 
calculations agree with those of Kayser \& Witt (1989), and
indicate that the maximum effective magnification of a uniform disk is
$\sim 1.4/\sqrt{\rho}$, where $\rho$ is the disk radius in units of the
Einstein radius of the lens.  The maximum effective magnification is slightly
greater for Lambert and for Lommel-Seeliger scattering.  Thus, an effective
magnification factor of $\sim 1.5 \times \sqrt{4000} \approx 100$ can be
achieved.}
As a result, as shown in the lightcurves below, the planet can perturb the
total flux by as much as $\sim 1\%$.

Blending of background starlight in crowded fields makes detecting
microlensing events more difficult.  For details, see GG00 and GCH03.  We
follow GCH03 and ignore blending, for with good seeing its effect is
negligible.

Note that the order in which the star and planet cross the caustic
matters a great deal for the detectability of the planet.  There are
two basic ways in which the planet-star system can be configured as it
crosses the caustic region -- planet leading star or planet trailing
star.  Since a planet will almost surely not be detected on ingress, as the
system enters the ICR, the favorable configuration for detecting a planet is
the planet trailing the star on egress, so that the star is not magnified
as the planet crosses the caustic.  If the planet is leading the star on
egress, the configuration is much less favorable for detecting the planet.

Finally, consider a future microlensing survey that uses a telescope large
enough to discern the ingress signature of a planet crossing the caustic.
Then, in a fraction of star-planet systems that cross fold-caustics, the system
could be in the favorable configuration for both caustic--crossings (i.e., with
the star outside the ICR when the planet crosses the caustic).
This is because the ICR--crossing--time ($\sim 3-4$ days) is
comparable to the semi-orbital period of a close-in extrasolar planet. For example,
consider a planet with orbital period $\sim 6$ days, twice the
ICR--crossing time of $\sim 3$ days.  In this case, there should be a
$\sim 50\%$ chance that the planet will be in the planet-leading
configuration on ingress and in the planet-trailing configuration on
egress after having traversed half an orbit (and an equal chance of
being unfavorably oriented both in ingress and egress); and so
a planet \emph{could} be detected on both ingress and egress.
Clearly, the actual likelihood of catching the same planet on both ingress
and egress crossings depends on the poorly known
distribution of orbital radii for both the planets and for the binary
lenses, but it is unlikely that the probability is negligibly small.
While the coincidence between the orbital and intra-caustic-region-crossing
timescales is interesting, we note that, in practice, a planet is unlikely to be 
detected on ingress -- unless a deep future survey is devoted to blind
monitoring of stars for lensing at $\sim$ hour time--resolution.

\section{Detectability}
\label{detectability}
\begin{figure}[ht]
\plotone{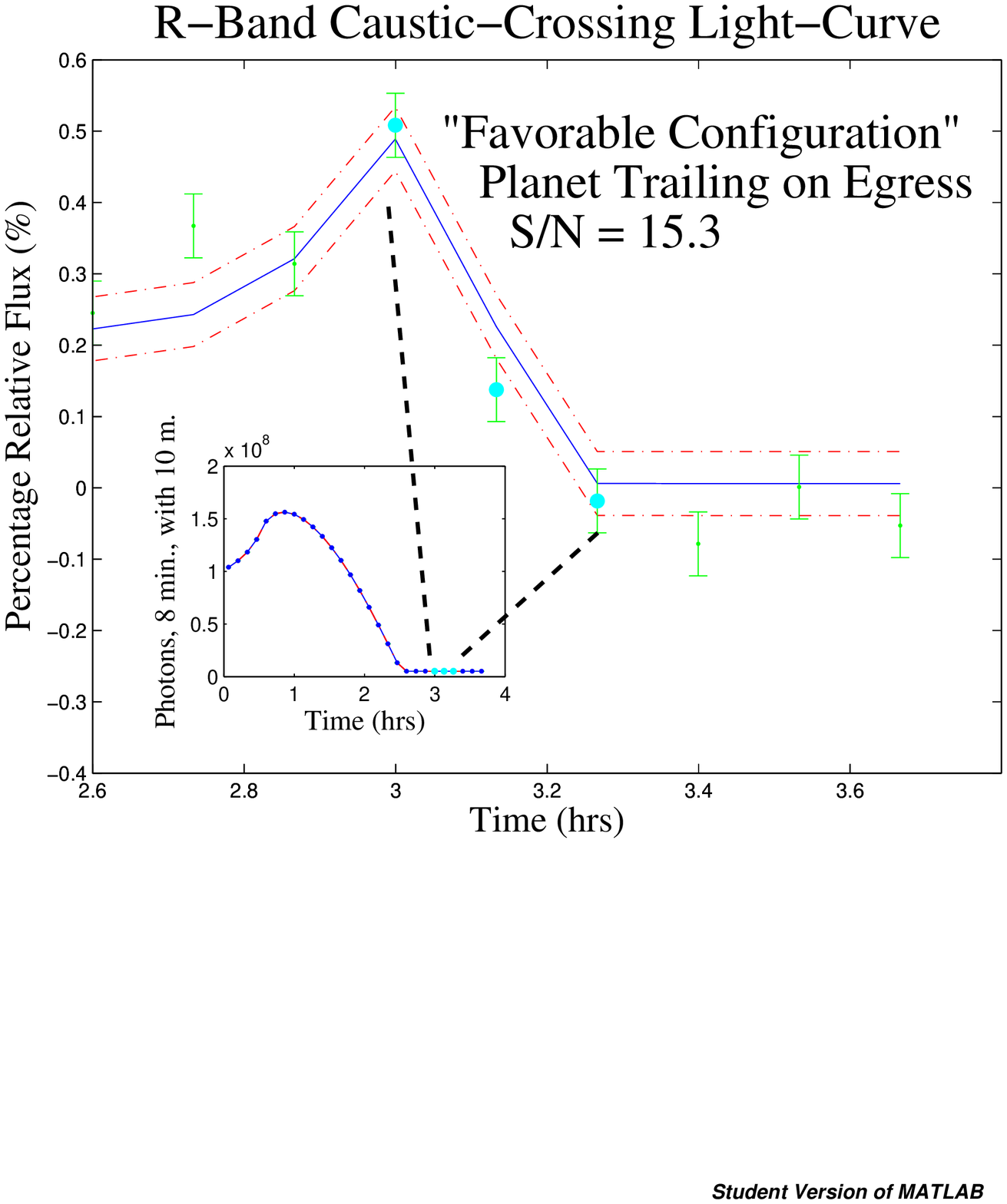}
\plotone{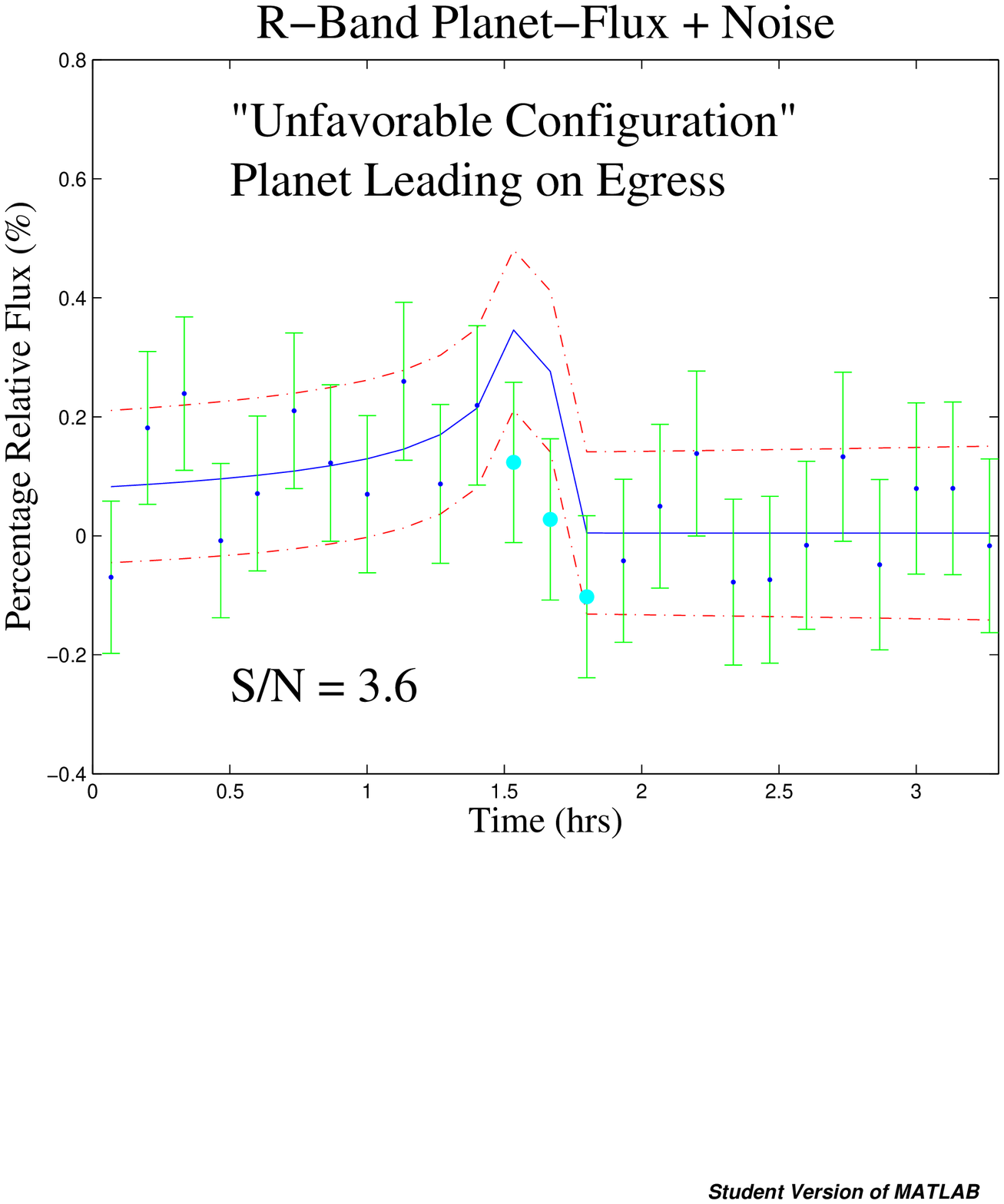}
\caption{Planetary lensing light--curves on egress in the $R$-band,
assuming Lommel--Seeliger reflection off a planet similar to HD20y9458b
and with Jupiter's albedo, expressed as percentage change in the total
flux caused by the presence of the planet.  Solid (blue) curves show
theoretical light--curves; dashed (red) curves show 1-$\sigma$ errors;
solid (blue) dots show random mock data (theoretical light--curve plus
noise) with 1-$\sigma$ error bars; large (cyan) dots denote times when
the planet's surface intersects the caustic. {\em Top panel:}
Favorable (planet--trailing) orientation for detection of the planet
on egress.  The inset shows the entire caustic-crossing lightcurve for
both the star and the planet, with the ordinate showing total photons
collected in 8 minute observations with a 10~m. telescope; the big
plot is a zoom-in of the tail end of the inset.  Dashed lines indicate
the relationship between the inset and the parent image.  An optimal
phase angle of $\phi=10\degr$ was assumed. {\em Bottom Panel:} Unfavorable
(planet--leading) orientation on egress, with the optimal value of
$\phi=-45^\circ$.}
\label{fig:tailhead} 
\end{figure}

Detecting the presence of a planet is, of course, challenging, since
even when the planet is on the caustic, its flux is a small fraction
($\lsim 1\%$) of even the un--magnified flux from the star.  As an
example, in the inset in Figure~\ref{fig:tailhead}, we present a model
$R$--band light--curve for a 10m telescope, showing first the star, and
then the planet exiting the ICR in the favorable configuration for detecting
the planet.  The planet is modeled with Jupiter's albedo (described in
\S \ref{prspectra} below), corresponding to $A\approx 0.45$. The solid
dots show simulated data points.  The broad peak between 0--2
hrs results from the star crossing the caustic.  The three large
dots at 3.0--3.3 hrs correspond to the planet
crossing the caustic.  On this scale, the magnified planetary flux is invisible
against the un--magnified star-flux.  Nevertheless, we next show that,
as we suggest in \S~\ref{preview} above, with the current generation
of 10m telescopes, it is possible to detect a planet when the star is
outside the ICR (but not when the star is inside the ICR).

In Figure~\ref{fig:tailhead}, the top panel shows the tail of the
light--curve (after the star has exited the ICR) for the planet's
caustic--crossing egress at $\phi=10\degr$ (i.e. a zoom--in version of
the planet signal from the inset). The bottom panel in this
figure shows a random realization of the flux from the planet for the
planet's caustic--crossing egress at $\phi=-45\degr$ (i.e. in the
unfavorable orientation).  In both panels, we show error--bars
corresponding to the $\sqrt{N}$ shot noise from the total
photon--flux.  (We ignore instrumental noise because shot noise
will dominate for bright bulge stars.)
We sum the signal and the $\sqrt{N}$ photon noise over five 8--minute
integrations around the planetary caustic crossing. In the favorable
orientation (the top panel), we find that the planet is detectable
with a total $S/N \sim 15.3$ while in the unfavorable orientation, the
planet is essentially undetectable ($S/N \sim 3.6$).  

The relationship between the signal-to-noise of detection and the
phase angle is summarized in Figure~\ref{fig:snvsphase} below.
The planet--flux/star--flux ratio is maximized when the planet is in
the ``full moon'' phase ($\phi \sim 0\degr$).  When the star is
outside the ICR, therefore, the planet's detectability is maximized
for low phase angles.  Phase $\phi\approx 10\degr$ is optimal (the
star intersects a fraction of the planet's surface for
$\phi\lsim 8\degr$, leading to a rapid decrease in the S/N ratio for
still smaller phase angles).  When the star is inside the ICR, however,
there is a more delicate balance.  Since magnification in the ICR
decreases with distance from the caustic, the planet's detectability
is improved when the the projected impact parameter $b$ is large, which
happens for $\phi \sim \pm 90\degr$.  These two competing
factors (planet--flux and star--flux) balance to maximize the
planet/star flux ratio at about $|\phi|\sim45\degr$.

The reader can estimate from Figure~\ref{fig:snvsphase} what fraction of
the orbit will yield acceptable signal--to--noise.  The $S/N$ of detection
exceeds 5 for approximately $90\degr$ of phase, or $1/4$ of the orbit. 
Since we are inquiring what will be possible in good viewing conditions,
we hereafter will consider only the favorable orientation ($\phi=10\degr$).

Note that, for most phases, the $S/N$ of detection is decreased if the plane of
the planet's orbit is inclined less than $90\degr$, but is unaffected if the
plane of the orbit is not normal to the caustic.

\begin{figure}[ht]
\plotone{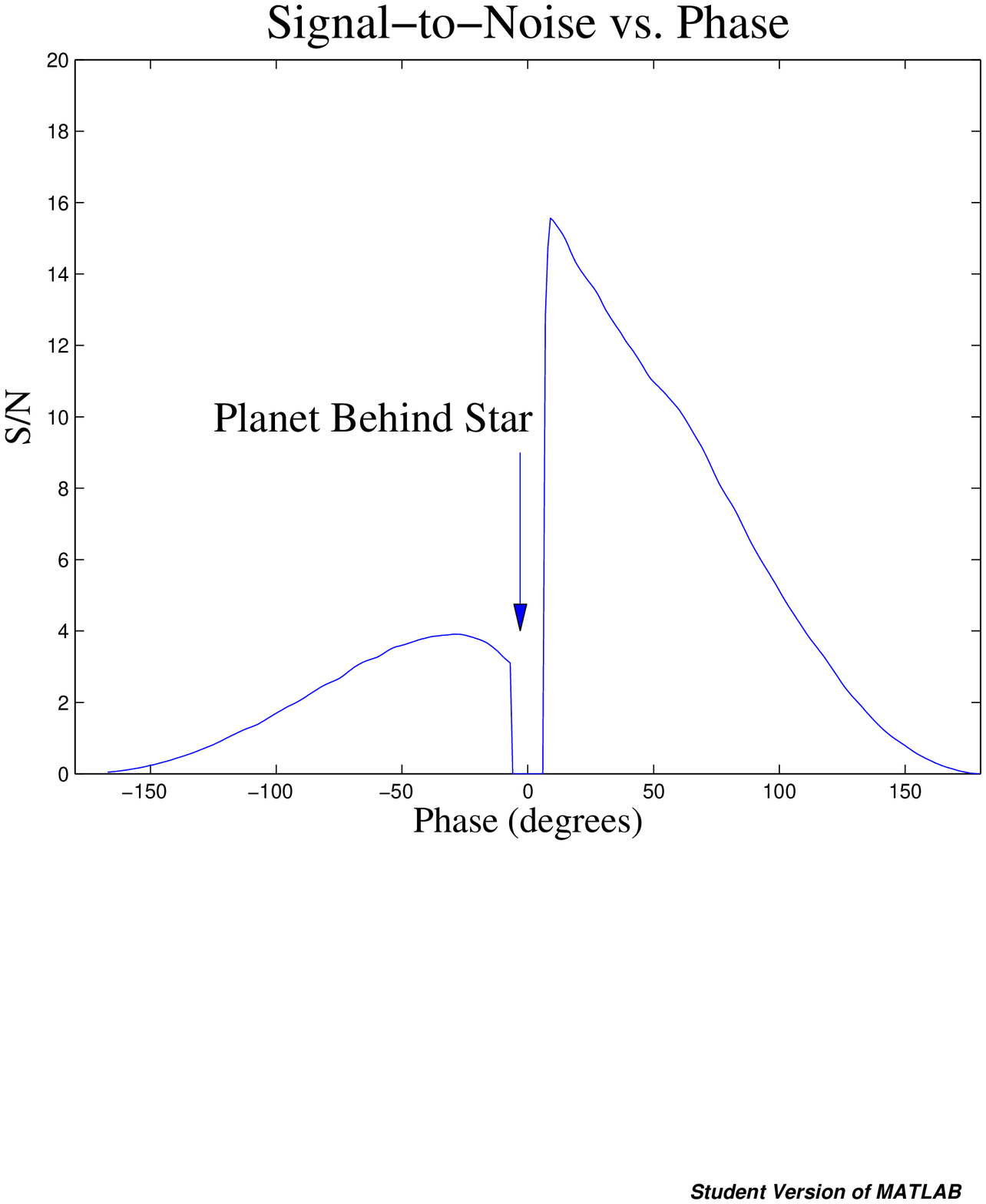}
\caption{This figure shows the dependence of $S/N$ on phase angle for $R$-band
detection of a planet in a clone of the HD209458 system at 8~kpc, lensed by
binary $0.3~M_\sun$ stars at 6~kpc, in 8-minute integrations with a 10~m
telescope.  The plane of the planet's orbit is assumed to be inclined
$90\degr$ to the line-of-sight (edge--on) and normal to the fold-caustic at the point of
crossing.  Typical proper-motion of lens and source (e.g. GCH03) is assumed.
Since the plane of orbit is at inclination $90\degr$, the planet disappears
behind the star for phase angles $|\phi|\lsim 8\degr$ and detectability drops
to zero.}
\label{fig:snvsphase}
\end{figure}

\section{Planetary Reflection Spectra}
\label{prspectra}
HD209458 is one of a relatively small number of stars confirmed to
have a transiting companion (HD209458b).
By carefully comparing the spectrum of this star during a planetary
transit against its spectrum outside transit, Charbonneau et
al. (2002) measured how the opacity of the transiting planet's
atmosphere varies with wavelength, and inferred the presence of sodium
in the atmosphere.  Performing a similar analysis, Vidal-Madjar et
al. (2003) claim to find an extended hydrogen Ly$\alpha$-emitting
envelope surrounding the planet.

We here investigate the prospects of analogously observing, instead of
a transmission spectrum during a transit event, a reflection spectrum
during a caustic--crossing event.
Although near--future ground--based coronographs 
(such as Lyot\footnote{http://lyot.org}) 
and more distant future space projects, such as the Terrestrial Planet
Finder\footnote{http://planetquest.jpl.nasa.gov/TPF/tpf$\_$index.html},
will be able to probe the spectra of planets around nearby stars
using coronographs or nulling interferometers (e.g., Kutchner \&
Traub, 2002), we are not aware of any other ways at this time to study the
reflection spectra of extrasolar planets.  We emphasize that the technique
we present in this paper is not in competition with current coronographic work,
but is rather in several senses complementary (for more information on
coronographic techniques, see, e.g., Oppenheimer et al. 2001).  First,
current and near--future coronographic studies are not sensitive to close--in
planets, because these planets lack sufficient angular separation from their
host--stars, but these are precisely the planets that are most readily seen in
the source--plane of a microlensing event.  Second, the population of planets
available to microlensing is in the Galactic bulge, or at a distance
$\sim8{\rm~kpc}$, and is therefore complementary to the nearby
population of planets that will eventually be available to the other
methods just mentioned.

As a first step toward modeling the
reflection spectrum of an extrasolar planet around a solar-type star,
we adopt the reflection spectra of the Jovian planets in our solar
system, because these are the only gas-giant planets whose wavelength-dependent
albedos have ever been measured.
Atmospheric conditions, and hence reflection spectra, of hot Jupiters
(extrasolar giant planets with short orbital periods) are likely to be
much different from those of Jupiter, Saturn, Uranus, and Neptune (for
detailed discussions of hot Jupiter atmosphere models, see, e.g.,
Sudarsky, Burrows, \& Hubeny, 2003; Burrows, Sudarsky, \& Hubeny,
2004; and Seager, Whitney, \& Sasselov, 2000).  However, given the
uncertainty and differences among published atmospheric models of
extrasolar giant planets, we prefer to base our calculations on the
unambiguously measured albedos of the solar--system gas giants. We
will then discuss (at the end of \S~5 below) the expected differences
for the hot--Jupiter atmospheres, and identify features in the
theoretical spectra that could be detected at a similar significance.

\begin{figure}[ht]
\plotone{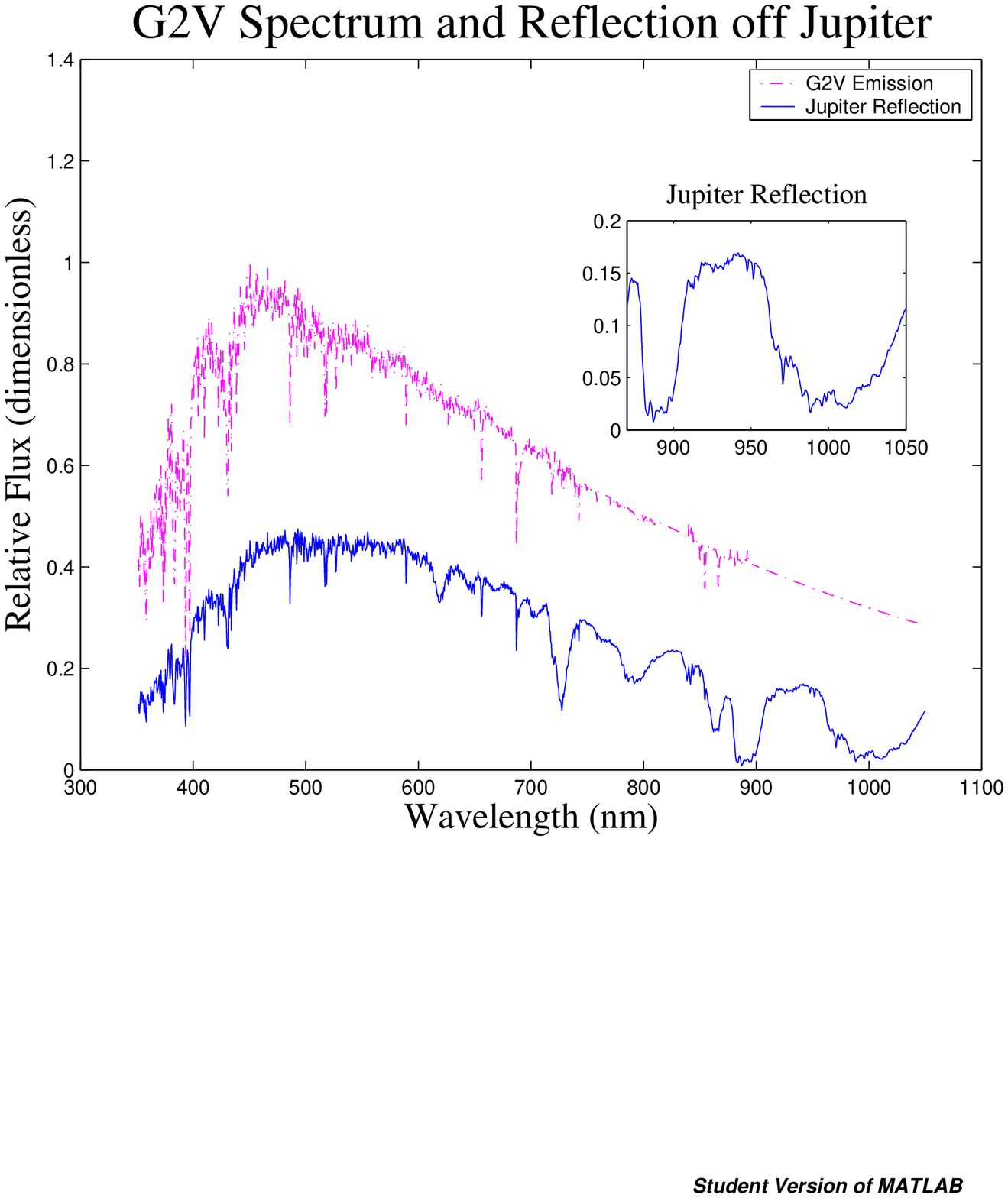}
\caption{Spectrum of a G2V star (top, dashed-dotted magenta curve) and reflection
spectrum of a G2V off a planet with Jupiter's albedo (bottom, blue curve).
Inset: zoom--in of Jupiter's reflection spectrum over the wavelength
range 870--1050nm.}
\label{fig:reflect} 
\end{figure}

\begin{figure}[ht]
\plotone{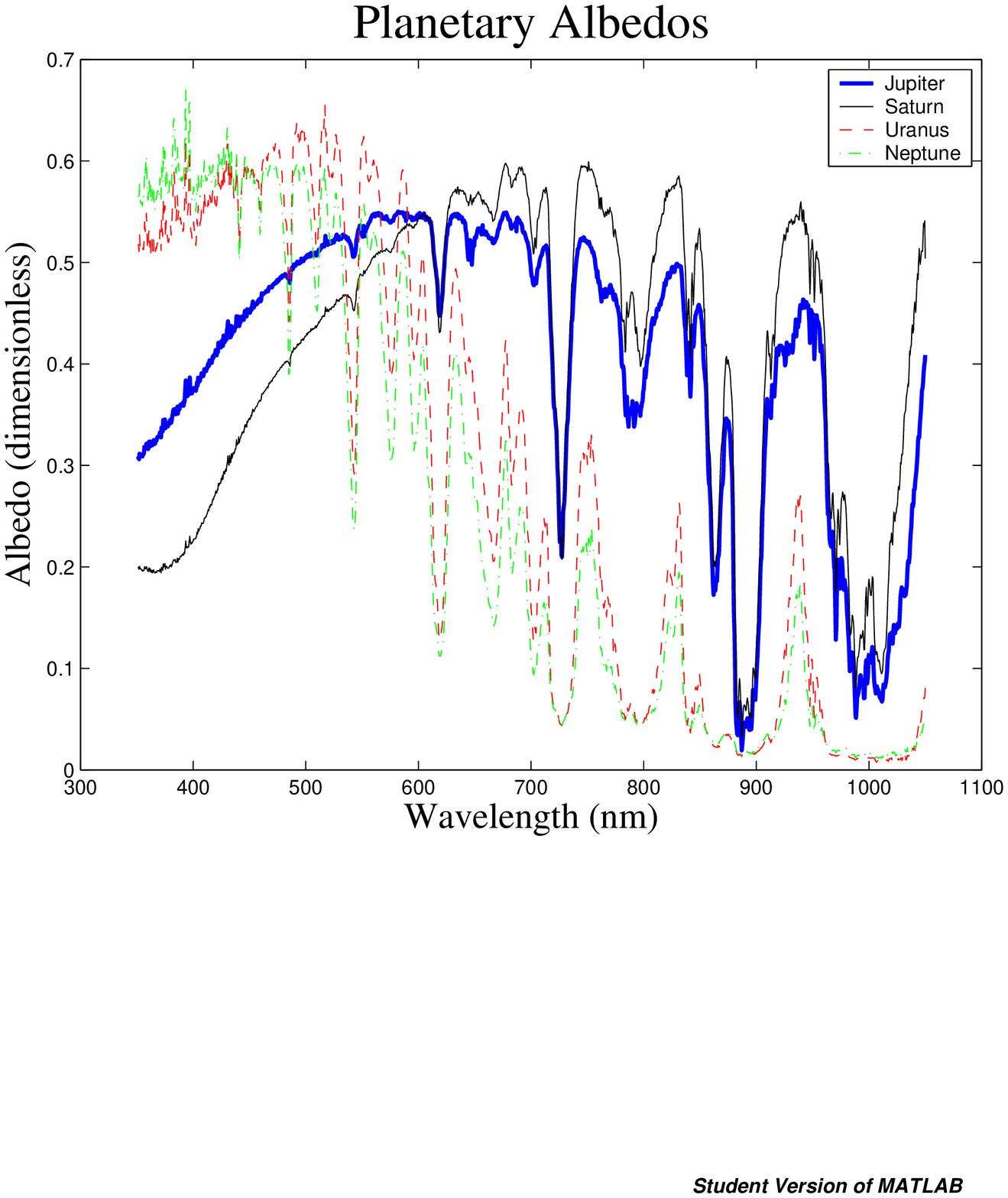}
\caption{Albedos of Jupiter (thick blue curve), Saturn (thin black curve), Uranus (dashed red), and
Neptune (dashed-dotted green), adopted from Karkoschka (1994).}
\label{fig:albedos} 
\end{figure}

To obtain our desired reflection spectra, we need the spectrum of a G2V star,
and the albedos of the gas giants in our solar system (with albedo defined as
the ratio of reflected flux to incident flux).  We obtained an
incomplete G2V spectrum from Greg Bothun's
webpage\footnote{\textsl{http://zebu.uoregon.edu/spectrar.html}}, that
had data missing at wavelengths of strong atmospheric absorption.
Regions of missing data up to 1050 nm were filled in with a best-fit
$T=6000~{\rm~K}$ blackbody spectrum, and the spectrum was normalized
to a peak value of unity for clarity of presentation (see
Fig.~\ref{fig:reflect}).  Planetary albedos are taken from Karkoschka
(1994), interpolated on a cubic spline (every 5nm) to the G2V
reference wavelengths, and are shown for the four gas giants in
Figure~\ref{fig:albedos}.  Reflection spectra (in arbitrary units),
then, are just the product of the albedo and the solar spectrum (shown
by the bottom curve in Fig.~\ref{fig:reflect}).

While a reflection spectrum of an extrasolar planet with a high
signal-to-noise ratio, covering a full range of wavelengths from the visible
into the near infrared (NIR), would be
ideal (see discussion in \S~\ref{specres} below), certain bands of the
visible and NIR spectrum provide more information about
chemical composition than others.  By comparing the reflected flux
from wavelength ranges where Jupiter's albedo is low with reflected
flux from comparable wavelength ranges where Jupiter's albedo is much
higher, we can infer the presence of those compounds responsible for
the low albedo.  It is clear from the bottom (blue) curve in
Figure~\ref{fig:reflect} that in a narrow band around 900nm
(880nm-905nm) and in a slightly wider band around 1000nm
(980nm-1030nm), Jupiter's albedo is quite low ($\sim 0.05$ and $\sim
0.1$, respectively); while in--between (920nm-950nm), its albedo is
much higher ($\sim 0.45$).  These troughs are caused by absorption by
methane in the Jovian atmosphere (Karkoschka 1994).  This stark
contrast in albedo between adjacent wavelength bands suggests a way to
search for, e.g., methane (or other elements or compounds that are expected, in
theoretical models for hot Jupiters, to cause features with a similar 
equivalent width; see discussion below) in the atmosphere of an 
extrasolar planet.

We note that extrasolar giant planets can orbit very close to their
host star (0.05 AU or less), but thermal emission from a planet would
nevertheless contribute negligibly to the reflected flux at
wavelengths $\sim1\mu$m even for a hot planet ($\sim 1500$K).

\section{Modeling Spectra During Caustic--Crossing}
\label{specres}
If the (unlensed) flux from the star has spectrum $F_*(\lambda)$ and
the planet has wavelength-dependent albedo $A(\lambda)$, then the flux
from the planet may be written
\begin{equation}
\label{Fp}
F_p(\lambda,t) = F_*(\lambda) A(\lambda) f(t),
\end{equation}
where the multiplicative function $f(t)$ depends on various
geometrical factors (the solid angle that the planet subtends from the
perspective of the star, whether the planet has moons or rings, how
far the planet is from the caustic, etc), and also on the reflectance
model.  The total observed flux, therefore, can be written (in the
favorable orientation, with the star un--magnified, as discussed
above) as
\begin{eqnarray}
\nonumber F_T(\lambda,t) & = & F_*(\lambda) + F_p(\lambda,t) \\
\label{Ft}
 & = & F_*(\lambda) ( 1 + A(\lambda)f(t) ).
\end{eqnarray}
The observables are $F_T$ and $F_*$.  The physically interesting
characteristics of the planetary system, however, are $A(\lambda)$ and
$f(t)$, and these may be solved for as
\begin{equation}
\label{Af}
A(\lambda)f(t) = \frac{F_T(\lambda,t)}{F_*(\lambda)} - 1 \equiv G(\lambda,t),
\end{equation}
where we define the function $G$ as the observable quantity
constructed on the right--hand side of equation~(\ref{Af}).  With perfect
data, the time-difference between the star's and the planet's caustic crossings
breaks the apparent degeneracy between $A$ and $f$ in the general solution,
\begin{eqnarray}
\label{A}
A(\lambda) & = & k_1 \mbox{exp}\left[\int\frac{\partial G/\partial \lambda}{G}d\lambda\right]\\
f(t) & = & k_2 \mbox{exp}\left[\int\frac{\partial G/\partial t}{G}dt \right],
\end{eqnarray}
(where $k_1$ and $k_2$ are constants of integration such that $Af = G$).

In practice, with data as noisy as can be expected with the current
generation of telescopes, it is impossible to separate $A$ from $f$,
and $A$ may be determined only given a model for $f$.  Still, it is
possible in principle to posit a model for $f$ (as outlined in
\S~\ref{model} above) and then to solve for $A(\lambda)$.  In this
case, since the signal--to--noise ratio for the detection of the
planet we find is only $\sim 15.3$, it is still only possible to split
the light into a few broad spectral bands, rather than into a resolved
spectrum.

In order to test the idea that we could look for the spectral
signature of a particular compound in the reflected light from a
distant extrasolar planet, we model a planet with Jupiter's
reflection-spectrum and scrutinize the model data for evidence of
methane.  In order to maximize signal--to--noise, we assume an egress
caustic--crossing with planetary phase $\phi=10\degr$.

\begin{figure}[ht]
\plotone{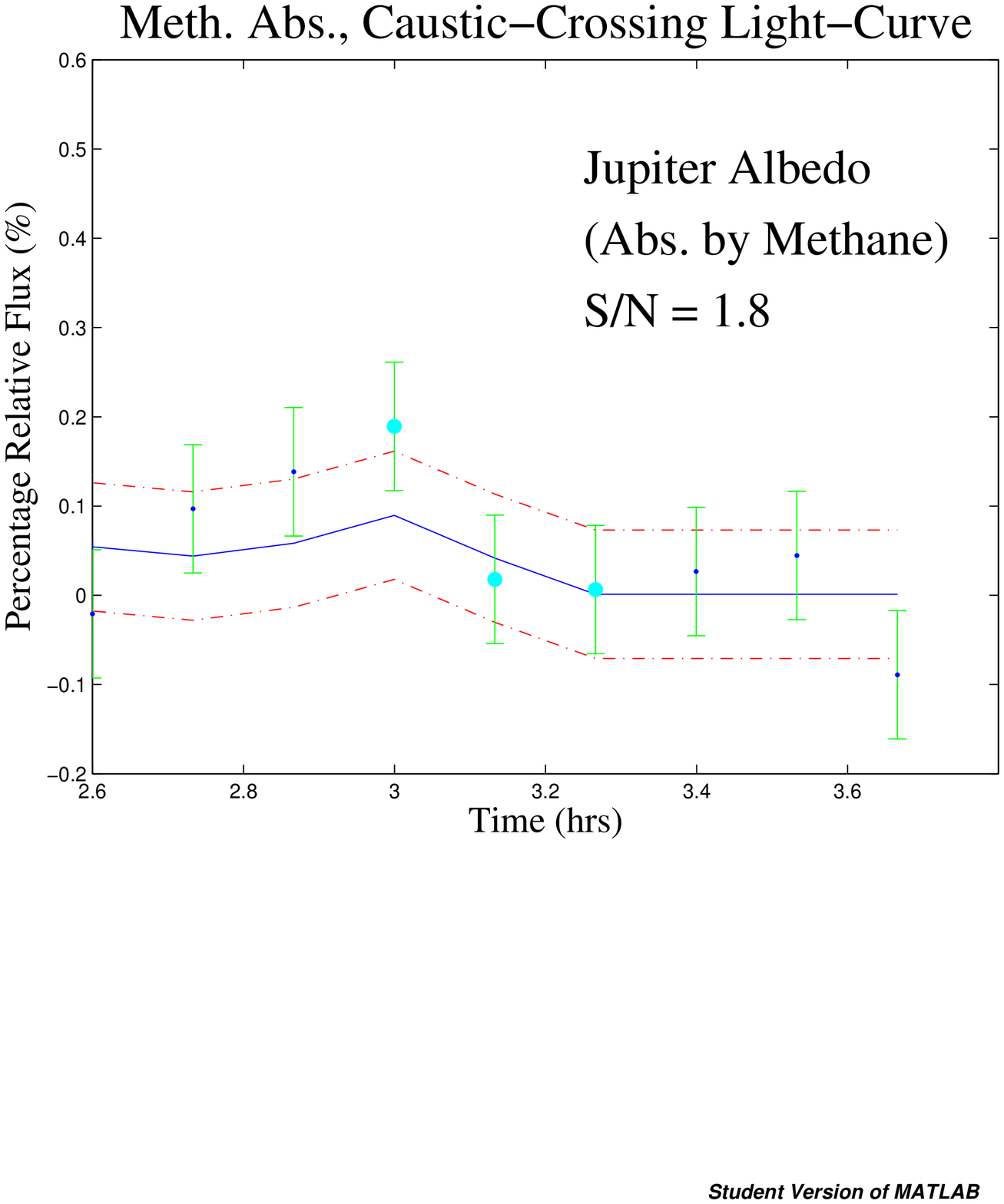}
\plotone{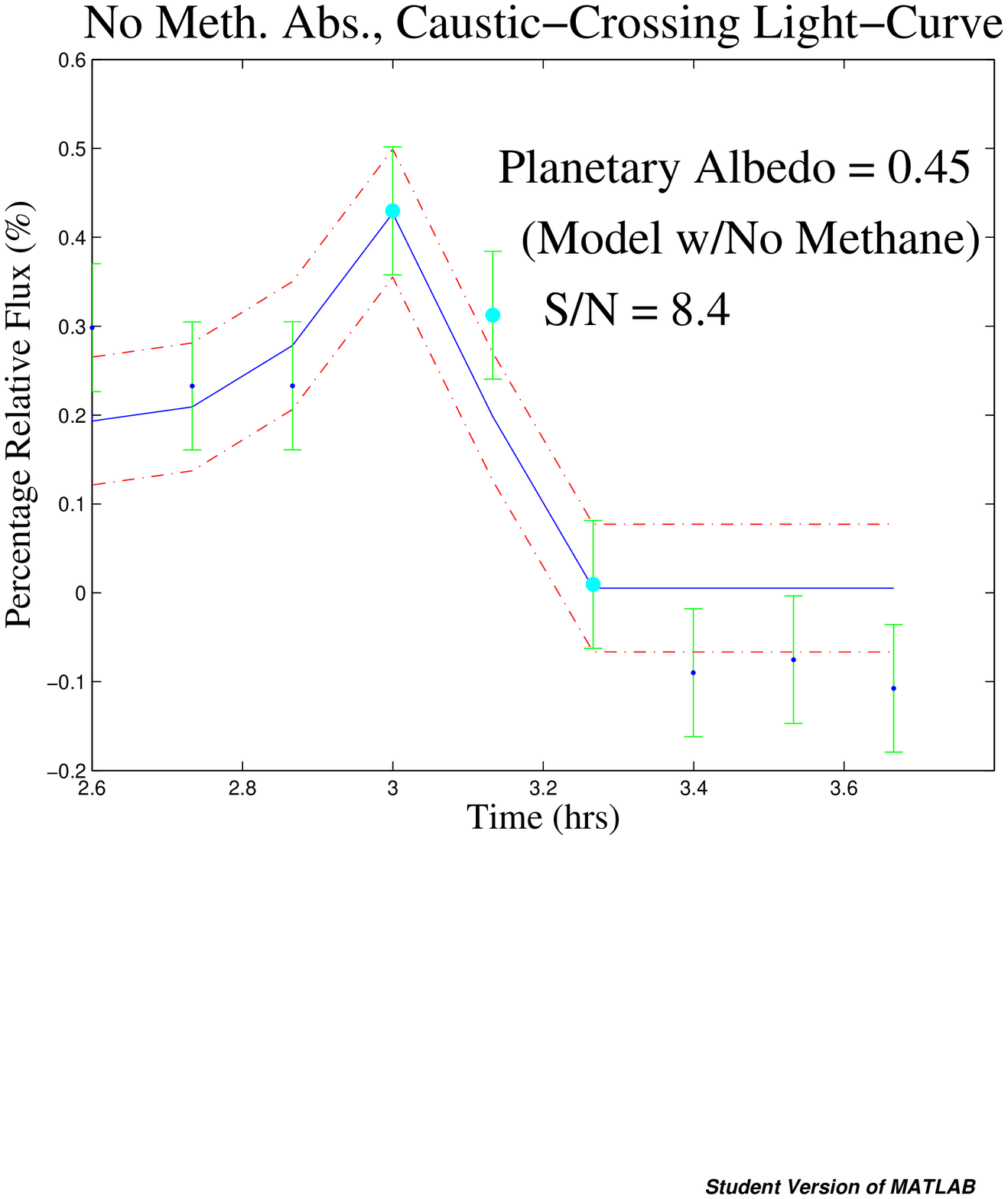}
\caption{Egress light--curves in the MBF band, which covers two deep
methane absorption features and includes light
from 880nm-905nm and from 980nm-1030nm.  The meaning of the symbols are as in
Figure~\ref{fig:tailhead}.  {\em Top panel:} Jupiter's low albedo is
adopted, which leads to a non--detection of the planet.  {\em Bottom
panel:} A constant albedo of $A=0.45$ is used showing what the light--curve
would look like if there were no methane present ($S/N = 8.4$).  This plot is quite
similar to the light--curve that would be obtained through a filter in a band
where there is low methane absorption and Jupiter's albedo is much higher ($\sim0.4-0.5$).}
\label{fig:methalb} 
\end{figure}

To search for signatures of methane, we construct a mock ``methane
band filter (hereafter ``MBF''), that allows complete transmission
from 880nm-905nm \emph{and} from 980nm-1030nm (the bands where
Jupiter's albedo is low because of methane, as discussed in \S
\ref{prspectra}) and zero transmission elsewhere (MBF is therefore a
``double top-hat'' filter).  Note that we do not necessarily mean
a physical filter here; we effectively assume that the flux in a
low-resolution spectrum can be binned and computed in these wavelength
ranges. A more realistic analysis would have to take into account the
additional instrumental noise in any physical implementation of such a
filter (such as read--out noise in the case of a spectrograph).  We
then compare the MBF light--curve of a model planet with Jupiter's
albedo to the MBF light--curve of a model planet with the methane
feature removed -- i.e., a model planet where the albedo is replaced
by a constant equal to Jupiter's mean albedo $\bar{A}=0.45$.  Figure
\ref{fig:methalb} shows this comparison: the top panel shows the MBF
light--curve for a planet with Jupiter's albedo (here, the planet is
detected at $S/N=1.8$, which counts as a non-detection); the bottom
panel shows the MBF light--curve for a planet with constant albedo
$A=0.45$ (here, the planet is detected at $S/N=8.4$).

In practice, the observational strategy would involve employing a
``high albedo filter'' (hereafter ``HAF'') that uses a region of the
spectrum that is relatively unaffected by methane and that is
comparable in width to the MBF filter (e.g. the adjacent 920nm-950nm
region, and/or other regions where Jupiter's albedo is high).  The
flux measured through the HAF filter would then be used to predict the
expected MBF flux according to the no--methane null--hypothesis.  In
practice, then, a \emph{non}--detection of the planet in the MBF band
together with a simultaneous \emph{detection} in the HAF band, would
be evidence for the presence of methane in the atmosphere of the
planet.
The $S/N$ is proportional to the square root of the number of photons
collected, and to the diameter of the telescope.  With a future 30 or
100m telescope, therefore, it would be possible to achieve a $S/N$ of
$\sim 25$ or $\sim 80$, respectively, in detecting the presence of
methane.

\section{Discussion}
\label{discussion}
Note that, strictly speaking, our $S/N$ calculations are for a space-based
observatory, because we do not include sky brightness.  Detailed data on
sky brightness are available in Lienert et al. (1998).  In R-band, the
contribution to total flux from the sky is small for good seeing (for seeing
$\sim0.75\arcsec$, the star is more than an order of magnitude brighter than
the sky within the aperture subtended by the star).  At $900$nm, the star is
still several times brighter than the sky for good seeing conditions, but by
$1\mu$m the sky is comparably bright to the star, which would increase the
noise in an observation by a factor of $\sim\sqrt{2}$ and would therefore
decrease the $S/N$ by the same factor.  For a 10m telescope, this still
indicates $S/N\sim6$ for good seeing conditions in the situation modeled above.
Note that if the plane of the planet's orbit is inclined less
than $90\degr$, then for nearly all phases the $S/N$ for detection of
the planet is reduced, for any spectral filter.
The ratio of the flux from the planet through the MBF to that through the
HAF, however, is independent of both inclination and phase.

Although some models of close--in extrasolar giant planets predict
significantly less methane than is present in Jupiter's visible
cloud-layer, this prediction is not universal.  Seager, Whitney,
\& Sasselov (2000), for example, present a model of 51 Peg b that has
spectroscopically significant methane--levels.  They point out,
however, that these methane features might be present only in the
coolest, least absorptive models.

Even if future models should converge upon the conclusion that the gaseous methane
content of hot Jupiters is very low, there are spectral features due to
other chemicals that are predicted to be present in the atmospheres of
close--in extrasolar giant planets, that are predicted to be comparably strong
to Jupiter's methane features.  A search for these other predicted features
would be analogous to the methods described above.
Sudarsky, Burrows, \& Hubeny (2003; SBH03) identify five classes of extrasolar giant
planets, ranging from class I (Jupiter--like) through classes IV ($a\sim0.1$AU)
and V ($a\sim0.05$).  A caveat introduced by SBH03 is that, for class IV and V
planets, the planet's spectrum redward of $\sim500$nm includes increasing levels
of thermal emission, and so it makes more sense to discuss the ``emergent spectrum''
rather than the reflection spectrum.  Their model emergent spectra for class IV planets
include several strong absorption features. In the visible, sodium ($\sim600$nm)
and potassium ($\sim800$nm) are predicted to induce absorption features
with a comparable equivalent width to the methane features we consider above;
in the NIR ($\sim1.4\mu$m), water, which is thought to condense too deep in
Jupiter to affect the cloud-top albedo, is predicted to cause an even
deeper (factor of $\sim100$) trough in the emergent spectrum.  This water
feature is at a wavelength where the Earth's sky is fairly bright (an order
of magnitude or more brighter than the star), which would make it difficult
to discern from ground-based observations but which should pose no difficulty
for a space-based telescope.  Table \ref{ta:eqwidths} summarizes the strengths
of the three absorption features predicted in SBH03 class IV planets described
above and compares them to the methane features previously considered.  If a planet
is detected with high $S/N$ in a HAF but is (un)detected in a filter centered on
a spectral feature with $S/N$ much less than the number quoted in the last column of
Table \ref{ta:eqwidths}, this would be evidence for the presence of the chemical
responsible for the feature.

The right-hand side of Figure \ref{fig:snvsphase} is summarized and
generalized in Equation~\ref{eq:SNcalc} below, which
gives a rough estimation of the expected signal--to--noise for 
detection of a planet whose orbit
is at inclination $90\degr$, in which the source is a Main-Sequence star.
The lensing configuration is taken to be the one described above.  $D$ is the
diameter of the telescope's aperture, $EW_f$ is the equivalent width of
the spectral filter used, and $EW_l$ is the equivalent
width of a spectral line or feature.
\begin{equation}
\label{eq:SNcalc}
\frac{S}{N} \sim \left( 16 - \frac{\phi}{9\degr}\right) \times
      \Theta(\bar{A}, D, EW_f, EW_l) \times \Psi(M_*, R_p, a)
\end{equation}
where $\Theta$ and $\Psi$ are the functions given below:
\begin{eqnarray}
\label{eq:Theta}
\Theta(\bar{A}, D, EW_f, EW_l) = \frac{\bar{A}}{0.45} \times
                                 \frac{D}{10{\rm~m}}
                                 \sqrt{\frac{EW_f}{150{\rm nm}}}
                                 \left(1-\frac{EW_l}{EW_f}\right) \\
\label{eq:Psi}
\Psi(M_*, R_p, a)  = 10^{-3} \left( \frac{2M_*}{M_\odot} -0.8 \right)
                            \left( \frac{R_p}{R_J} \right)^{3/2} 
                            \left( \frac{a}{1~{\rm AU}}\right)^{-2}.
\end{eqnarray}
Observe that both $\Theta$ and $\Psi$ are unity in the case of an $R$-band
observation through a 10 m telescope of the planetary system considered above.

Equation \ref{eq:SNcalc} slightly over-predicts $S/N$ for filters on the red
side of $R$-band and slightly under-predicts $S/N$ for filters on the blue
side; furthermore, as above, it does not include sky-brightness, which is
particularly important redward of $1\mu$m.

\begin{deluxetable}{lcrrr}
\tablewidth{240pt}
\tablecolumns{4}
\small
\tablecaption{Absorption Features\tablenotemark{a}}
\tablehead{\colhead{Spectral} & \colhead{SBH03} & \colhead{Line}   & \colhead{Eq.}   & \colhead{$S/N$} \\
           \colhead{Feature}  & \colhead{Class} & \colhead{Center} & \colhead{Width} & \colhead{Ratio}       }
\startdata
Sodium           &  IV          & $\sim600$nm    & $\sim80$nm  & 14  \\
Potassium        &  IV          & $\sim780$nm    & $\sim20$nm  &  7  \\ 
Methane          &  I           & $\sim990$nm    & $\sim20$nm  &  5  \\
Methane          &  I           & $\sim1.00\mu$m & $\sim40$nm  &  7  \\
Water            &  IV          & $\sim1.40\mu$m & $\sim200$nm & N/A \\
\enddata
\label{ta:eqwidths}
\vspace{-0.4cm}
\tablenotetext{a}{This table shows the line-center and equivalent width
for each of 5 spectral features.  Three of these features, the Sodium (Na),
Potassium (K), and Water features, are expected in planets classified by
SBH03 as class IV ($a\sim0.1$AU).  The other two features are the methane
features considered in detail in this paper, with data taken from Jupiter's
reflection spectrum, available in Karkoschka (1994).
The last column shows the predicted $S/N$ ratio given the absence of the
chemical, and for Na and K it is computed from Equation \ref{eq:SNcalc}
(using $\bar{A}=0.35$ and $EW_l=0$), while for the Methane features it is
computed from our simulations.
This ratio is not applicable to the case of water, because the sky is
too bright for ground-based observations at this wavelength.}
\end{deluxetable}

\section{Conclusions}
\label{conc}

In the Galactic bulge there is a large number of stars and,
presumably, a comparably large number of planets.  With current and
future microlensing surveys in the direction of the bulge, we expect
that some solar systems will cross the fold-caustics of binary lenses.
Unfortunately, although in such events there will be two
caustic--crossings, it appears that current technology will only allow
for detection of a planet orbiting the source-star during the egress
caustic--crossing -- and furthermore only when the star-planet system is in
the favorable configuration.  Still, with its expected mean albedo, the planet
should reflect enough light that, in the case we consider, it should be
detectable for roughly $1/4$ of its orbit with a 10m telescope.

Our results suggests that the strategy outlined by GG00 and Lewis \& Ibata
should be viable: each time a bulge star is seen to cross a fold-caustic into
the ICR, the egress event should be closely monitored in order to
detect a planet in the trailing (favorable) configuration, should such
a favorable orientation occur.  If $10\%$ of bulge stars have hot
Jupiter companions, then, since a quarter of planet-star systems will be have
appropriate in
the planet-trailing configuration on egress, $\sim2-3\%$ of bulge stars
that cross fold-caustics will be seen, under close monitoring (in
$\sim$ 8--min integrations) during the egress crossing, to have
planetary companions.

If such planets are detected, it will be possible, in principle, to
determine various properties of the planet, including physical
(reflectance model, phase, angular orientation relative to the
caustic, presence of moons or rings; see GCH03) and chemical
characteristics (the presence of specific constituent compounds of the
atmosphere, as suggested by our results). Since the expected
perturbations to an observed light--curve from the physical
characteristics are either small (moons, rings, angular orientation)
or degenerate with other effects on the total brightness of the
planet, such as the planet's albedo or the solid angle it subtends
from the perspective of its star (reflectance model, phase), it will
be difficult in practice to determine these physical characteristics.
For example, if we were to observe the egress caustic-crossing light--curve of
a planet in the bulge that has rings around it, is at illumination phase
$\phi=45^\circ$, and obeys Lommel-Seeliger reflection, we would most likely
not be able to infer the presence of the rings, the phase, or the
nonuniform reflectance because the data could be fit
equally well (within error bars) by a best-fit $\phi=0^\circ$ model with no 
rings (at $\phi=0^\circ$, a Lommel-Seeliger planet is uniformly illuminated).
The expected perturbations from some atmospheric compounds, however,
are much greater (a factor of $\sim5$ or more) and do not suffer from
analogous geometrical degeneracies.  

In our example, using 8--minute observations on a 10m telescope, we
found that the presence of methane could be inferred from a
non--detection of the planet in the strong broad methane absorption
band at $\approx 0.9\mu$m, accompanied by a $S/N\sim 10$ detection in
adjacent bands.  Observations such as the ones described in this paper
will provide a crucial constraint on models of roaster atmospheres.
Then, in turn, as more accurate atmosphere models become available,
this $S/N$ could improved by fitting the data to
spectral templates with free parameters corresponding to variable
compositions.  Future generations of large telescopes might therefore be 
able to probe the composition of the atmospheres of distant extrasolar
planets.

\vspace{\baselineskip} 
We thank Scott Gaudi for extremely helpful discussions and criticism,
Erich Karkoschka for providing planetary albedo data, and Sara Seager
for commentary on theoretical hot Jupiter atmospheres.  We also thank
our referee, David Graff, for suggesting many improvements upon the
original version of this paper.

\section*{Appendix}
\label{appendix}

In this Appendix, we present the specifics of our model of the planet,
including details regarding the three reflectance models we consider.
\begin{figure}[ht]
\plottwo{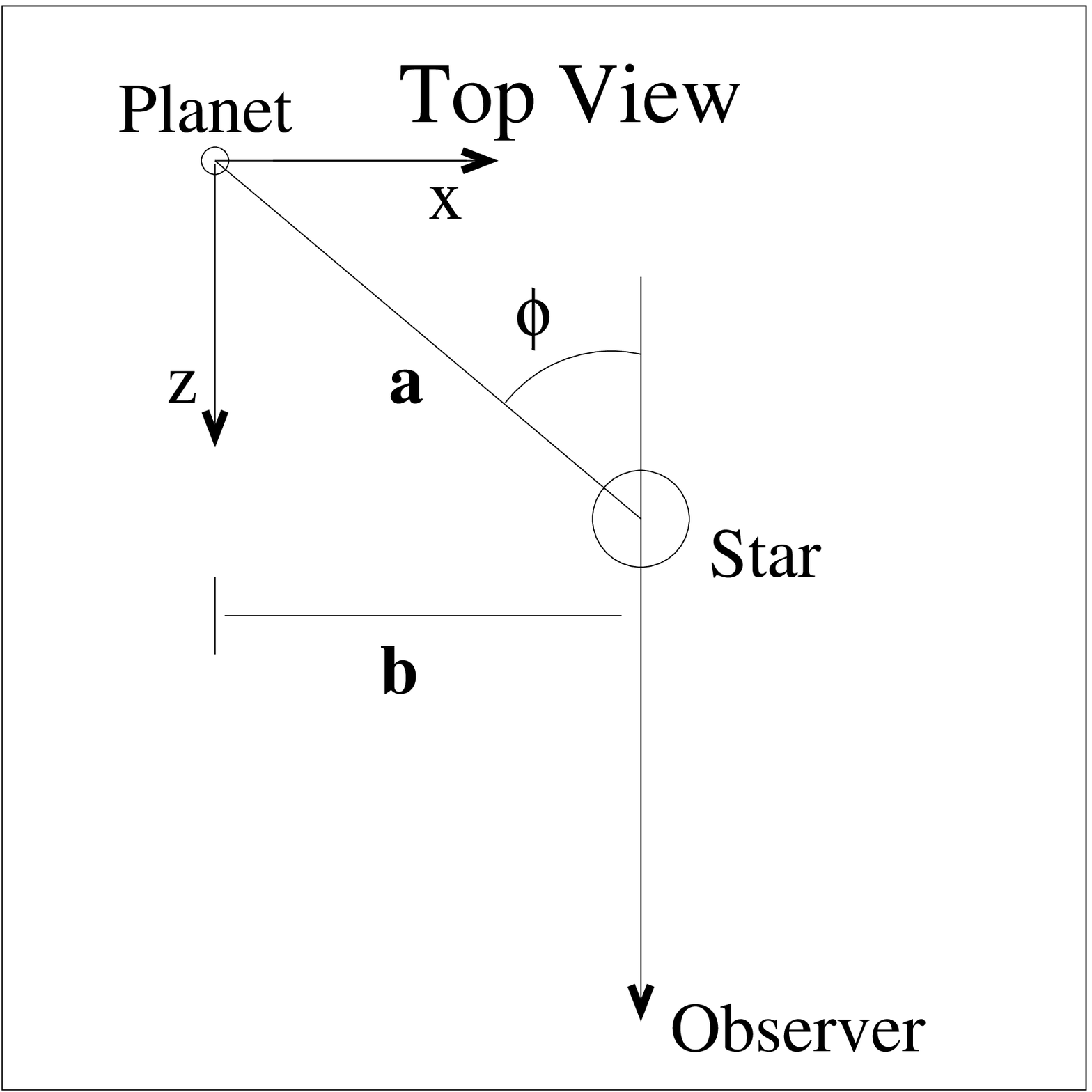}{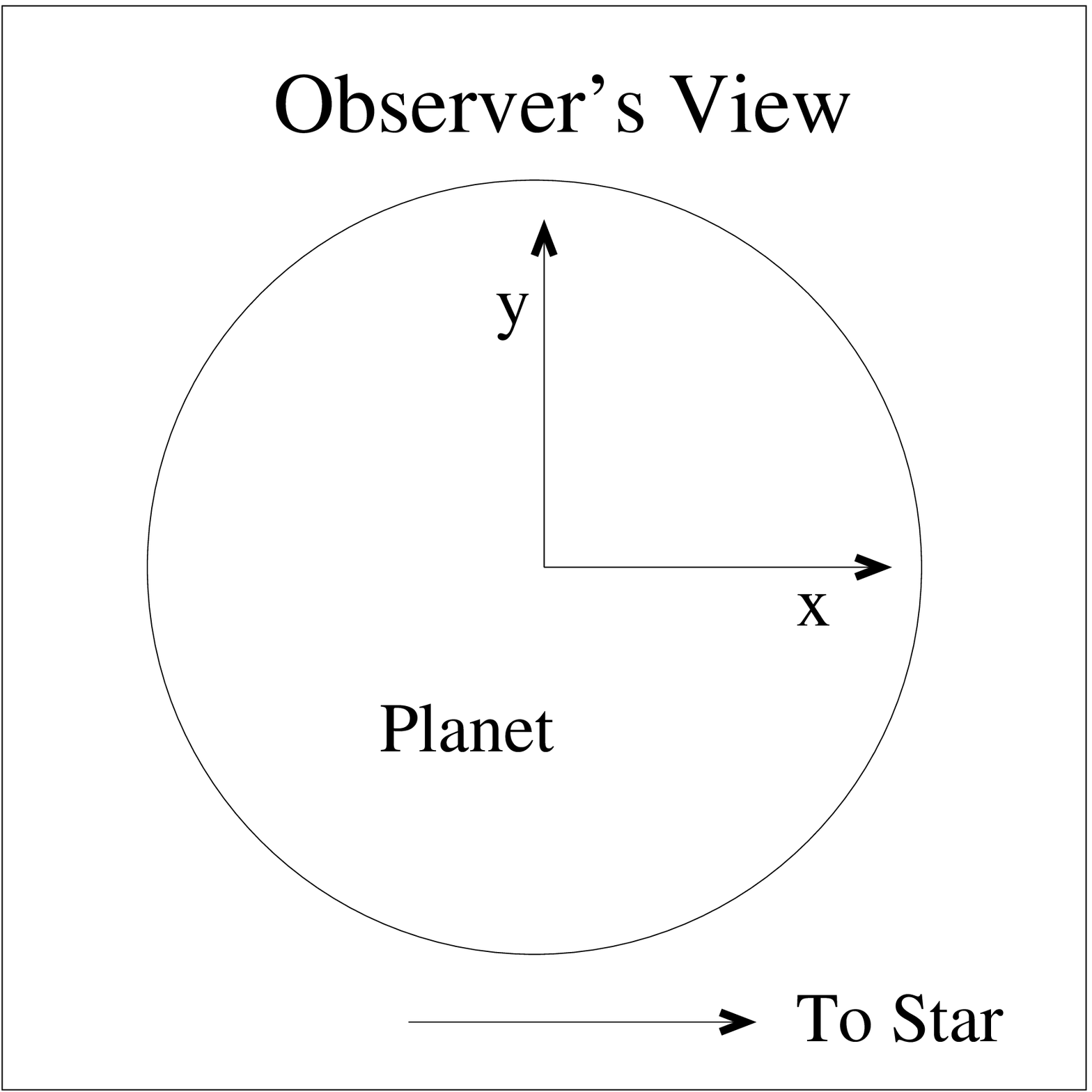}
\caption{{\em Left panel:} Top view (``God's eye view'') of the
viewing geometry, showing the phase angle ($\phi$), the orbital radius
($a$), and the projected impact parameter ($b = a \mathrm{sin}\phi$).
{\em Right panel:} Observer's view of the planet.}
\label{fig:viewgeom} 
\end{figure}

The viewing geometry of a planet--star binary can be described in a
three--dimensional coordinate system, centered on the planet, as
illustrated in Figure~\ref{fig:viewgeom}.  The $z$--axis is defined to
point toward the observer; the $x$--axis is in the direction from the
planet to the star, as projected on the sky from the perspective of
the observer; and the $y$--axis is defined by the $x$ and $z$-axes and
the usual right--hand rule.  The phase is as defined in Section \ref{pssystem}
given by the angle.

The reflectance models considered are the following:
\noindent
{\it Uniform reflection:} the planet has uniform surface--brightness
as seen from the observer, regardless of phase.
{\it Lambert reflection:} the surface--brightness of a patch of
projected surface is proportional to the cosine of the angle between
the incident radiation and the surface--normal vector: $B \propto
\cos(\theta_{\rm{ill.}})$.  Let $\hat{x}$, $\hat{y}$, and $\hat{z}$ be the
dimensionless coordinates on the planetary surface ($\hat{x} = x/R_p$;
$\hat{y} = y/R_p$, and $\hat{z} = z/R_p$, where $R_p$ is the planet's radius).
The unit vector to the star is $\mathbf{s} = (\sin\phi,0,\cos\phi)$, and the
unit surface-normal vector is $\mathbf{N} = (\hat{x},\hat{y},\hat{z}) =
(\hat{x},\hat{y},\sqrt{1-(\hat{x}^2 + \hat{y}^2)})$, so the desired cosine
is given by $\mathbf{s} \cdot \mathbf{N} = \hat{x} \sin\phi +
\sqrt{1- \hat{x}^2 - \hat{y}^2}\cos\phi$.  A failing of the Lambert
reflectance model is that in the ``full-moon'' phase, the specific intensity
from the edge of the projected disk drops to zero, in conflict with the
appearance of the Moon and other planets in our Solar system.
{\it Lommel--Seeliger reflection} is a phenomenological model,
designed to reproduce the reflectance of the Moon, that also mimics
well the appearances of a number of other bodies in the Solar System.
Neither the Lambert nor the Lommel--Seeliger model -- and certainly
not the uniform model -- can capture in detail the appearance of a
patch of planetary or Lunar surface at high resolution; but the
Lommel-Seeliger model in particular is successful at reproducing at
low resolution the whole planetary disk.  The surface--brightness of a
patch of projected surface in the Lommel-Seeliger model is
proportional to the cosine of the illumination angle, and inversely
proportional to the sum of the cosines of the illumination angle and
the viewing angle: $B \propto
\cos(\theta_{\rm{ill.}})/[\cos(\theta_{\rm{ill.}}) +
\cos(\theta_{\rm{view}})]$.  The cosine of the viewing angle is the
dot product of $\mathbf{N}$ with the unit vector to the observer,
$\mathbf{z} = (0,0,1)$, or $\mathbf{N} \cdot \mathbf{z} = \sqrt{1 -
(\hat{x}^2 + \hat{y}^2)}$.

Within our coordinate system, the un--magnified flux from a patch of
surface at projected coordinates $(\hat{x},\hat{y})$ can be represented as
follows
\begin{equation}
dF = K P \frac{B(\hat{x}, \hat{y})}{4 \pi r^2} d\hat{x} \spa d\hat{y}.
\end{equation}
Here $P$ is the total incident stellar power that the planet reflects,
or $L_* A (\pi R_p^2)/4 \pi a^2$, where $L_*$ is the luminosity of the
star; $A$ and $a$ are the planet's albedo and orbital
semi-major axis, respectively; $B(\hat{x}, \hat{y})$ gives the spatial
dependence of the apparent brightness of the planet, and depends on
the reflectance model; $r$ is the distance of the observer from the
system; and $K$ is an overall scale-factor so that the total reflected
light equals the total intercepted light times the albedo $A$.  What
remain to be given, then, are $K$ and $B$ for each reflectance model.

The resulting constants and formulae are:
%
%
\begin{eqnarray}
\nonumber K_U    & = & 2/ \pi \\
\nonumber K_L    & = & 4/ \pi \\
K_{LS} & = & 1.556,
\end{eqnarray}
and
\begin{eqnarray}
\nonumber B_U(\hat{x},\hat{y}) & = & 1 \\
\nonumber B_L(\hat{x},\hat{y}) & = & \hat{x}\sin\phi + \sqrt{1-(\hat{x}^2+\hat{y}^2)}\cos\phi \\
B_{LS}(\hat{x},\hat{y}) & = & \frac{\hat{x}\sin\phi + \sqrt{1-(\hat{x}^2+\hat{y}^2)}\cos\phi}{\hat{x}\sin\phi + \sqrt{1-(\hat{x}^2+\hat{y}^2)}(\cos\phi + 1)},
\end{eqnarray}
where the $U$, $L$, and $LS$ subscripts refer to uniform, Lambert, and Lommel-Seeliger reflectance, respectively.

\end{document}